\documentclass[letterpaper,12pt]{article}
\usepackage[letterpaper, top=2cm, bottom=2.5cm, left=2cm, right=2cm]{geometry}

\parskip=\baselineskip
\renewcommand{\baselinestretch}{1.2}

\usepackage{amssymb,amsmath}
\usepackage[english]{babel}
\usepackage{latexsym}
\usepackage{hyperref}

\newcommand{\dd}{{\rm d}}

\newcommand{\comm}[2]{\left[#1,#2\right]}

\newcommand{\eq}{\begin{equation}}
\newcommand{\feq}{\end{equation}}
\newcommand{\eqn}{\begin{eqnarray}}
\newcommand{\feqn}{\end{eqnarray}}
\newcommand{\equ}{\begin{eqnarray*}}
\newcommand{\fequ}{\end{eqnarray*}}

\newcommand{\M}{{\cal M}}

\newcommand{\F}{{\cal F}}
\newcommand{\A}{{\cal A}}

\begin{document}
\begin{titlepage}
\begin{center}
$\phantom{1}$

\vspace{1.5cm}

{\textbf{\Large Gravity in the 3+1-Split Formalism II:\\}}
\vspace{.1cm}
{\textbf{\Large Self-Duality and the Emergence of the}}\\
{\textbf{\Large Gravitational Chern-Simons in the Boundary}}

\vspace{1cm}

{\large D.~S.~Mansi\footnote{\href{mailto:dmansi@physics.uoc.gr}{\tt dmansi@physics.uoc.gr}} and A.~C.~Petkou\footnote{\href{mailto:petkou@physics.uoc.gr}{\tt petkou@physics.uoc.gr}}}

\vspace{0.2cm}

{\it Department of Physics, University of Crete, GR-71003 Heraklion.}

\vspace{0.5cm}

{\large G.~Tagliabue\footnote{\href{mailto:giovanni.tagliabue@mi.infn.it}{\tt giovanni.tagliabue@mi.infn.it}}}

\vspace{0.2cm}

{\it Dipartimento di Fisica dell'Universit\`a di Milano, Via Celoria 16, I-20133 Milano\\
and INFN, Sezione di Milano, Via Celoria 16, I-20133 Milano.}
\end{center}

\vspace{.5cm}

\begin{abstract}
We study self-duality in the context of the  3+1-split formalism of gravity with non-zero cosmological constant.  Lorentzian self-dual configurations are conformally flat spacetimes and have boundary data determined by classical solutions of the three-dimensional gravitational Chern-Simons. For Euclidean self-dual configurations, the relationship between their boundary initial positions and initial velocity is also determined by the three-dimensional gravitational Chern-Simons. Our results imply that bulk self-dual configurations are holographically described by the gravitational Chern-Simons theory which can either viewed as a boundary generating functional  or as a boundary effective action. 
\end{abstract}

\end{titlepage}

\parskip=.1\baselineskip
\tableofcontents

\parskip=.3\baselineskip
\renewcommand{\baselinestretch}{1.2}

\section{Introduction}
In a previous work \cite{MPT1} we embarked into a detailed analysis of gravity with a non-zero cosmological constant in the 3+1-split formalism having in mind applications to holography and AdS$_4$/CFT$_3$. In \cite{MPT1} we setup the formalism, fixed notation and defined the various quantities of interest. We have then argued that holography can be viewed as an initial value problem at the boundary and we have shown that the various methods for the removal of divergences can be interpreted as suitable canonical transformations to define a well-posed Cauchy problem at the boundary. As an application of our ideas we have considered the holographic description of various black hole solutions. 

In this work we continue our analysis of gravity with non-vanishing cosmological constant in the 3+1-split formalism. Our aim is to study self-duality, both in Lorentzian and Euclidean signatures. The 3+1-split formalism \cite{LP1} is perfectly sui\-ted for such studies since it provides a clear definition of the ``electric" and ``magnetic" gravitational fields. Our results unveil the intriguing role the three-dimensional gravitational Chern-Simons theory plays in the holographic description of four-dimensional gravity  both for Euclidean as well as for Lorentzian signature. The notion of self-duality in gravity with non-vanishing cosmological constant is connected with the self-duality of the on-shell Weyl tensor. In the case of Lorentzian signature this implies the vanishing of the Weyl tensor and hence the bulk self-dual configurations are confomally flat metrics\footnote{Such configurations we first studied holographically in \cite{SS}.}. For such metrics the boundary data are fully determined by classical solutions of the three-di\-men\-sio\-nal gravitational Chern-Simons theory. In the case of Euclidean signature we can have non-trivial configurations with self-dual Weyl tensor. Their boundary data, namely {\it boundary initial position} and {\it boundary initial velocity} in the terminology of \cite{MPT1}, are related via the three-dimensional gravitational Chern-Simons functional. Explicitly, the {\it boundary initial velocity} is given by the Cotton tensor of the boundary vielbein, the latter being non-constrained. The standard holographic interpretation of our result is that the three-dimensional gravitational Chern-Simons theory is the {\it exact} boundary generating functional of connected diagrams for the CFT dual to bulk self-dual gravity configurations. However, we point out that our results indicate an intriguing alternative interpretation; the three-dimensional gravitational Chern-Simons functional is the leading order {\it effective action} of the boundary theory. 

In section 2 we briefly review the 3+1-split formalism of \cite{LP1,MPT1}. In Section 3 we show that the self-dual bulk configurations in the case of Lorentzian signature, namely the conformally flat metrics, have boundary data determined by the classical solutions of the three-dimensional Chern-Simons theory. In Section 4 we turn our attention to Euclidean signature. Firstly we discuss the standard notion of self-duality in the absence of cosmological constant, and show how known results are nicely reproduced in the 3+1-split formalism. Next we discuss self-duality in the presence of a non-zero cosmological constant.  In that case, the self-duality refers to the on-shell Weyl tensor. Again, known solutions arise in an economical way in the 3+1-split formalism. Finally, we show that the requirement of bulk self-duality reduces to a condition between the boundary data. Explicitly, the {\it boundary initial velocity} is given by the Cotton tensor of the boundary veilbein. This results shows that the holographic boundary generating functional for self-dual configurations is the gravitational Chern-Simons functional.  Nevetheless, our results motivate a modification of the holographic dictionary, in the spirit of \cite{dHP1,dHG,dHPP,dHPConf,Marolf},  by which the on-shell gravitational action for self-dual configurations can be interpreted as the leading term of the boundary effective action.  Two Appendices contain useful relations for the Weyl tensor and also a brief presentation of the fist-order formalism for Yang-Mills theories. 

\section{Resum\'e of the 3+1-split formalism in Lorentzian signature}
\label{resume}
In this section we briefly review the results of \cite{MPT1} concerning the 3+1-split formalism for gravity  in the presence of a cosmological constant. The starting point is the Einstein-Hilbert action
\eqn
S_{{\rm EH}}=-\frac1{32\pi G}\int_{\M}\epsilon_{abcd}\left[R^{ab}+\frac{\sigma_{\perp}}{2\ell^{2}}e^{a}\wedge e^{b}\right]\wedge e^{c}\wedge e^{d}\label{EH.action}
\feqn
where $\M$ is a Lorentzian manifold, the tangent metric is $\eta_{ab}={\rm diag}(\sigma_{\perp},+$, $+,-\sigma_\perp)$, where $\sigma_{\perp}=\pm1$, and the cosmological constant is related to the characteristic length $\ell$ by $\Lambda_{{\rm cosm.}}=-3\sigma_\perp/\ell^2$.  Hence $\M$ can be foliated by slices $\Sigma_{t}$ indexed by a function $t$ which is either a time coordinate if $\sigma_{\perp}=-1$ or a radial coordinate if $\sigma_{\perp}=+1$. Accordingly we split the vielbein and the spin connection as\footnote{Throughout this work, Latin indices run as $a,b,c...=0,1,2,3$ and Greek indices as $\alpha,\beta,\gamma,...=1,2,3$.} 
\eqn
\label{esplit}
e^0&=&N\dd t\,,\qquad e^\alpha=N^\alpha\dd t+\tilde e^\alpha\,,\\
\label{omegasplit} \omega^{0\alpha}&=&q^{0\alpha}\dd t+\sigma_{\perp}K^\alpha\,,\qquad\omega^{\alpha\beta}=-\epsilon^{\alpha\beta\gamma}\left(Q_\gamma\dd t+B_\gamma\right)\,,
\feqn
where $K^\alpha$ and $B^\alpha$ are the ``electric" and ``magnetic" fields respectively. The Einstein-Hilbert action \eqref{EH.action} with the addition of the boundary Gibbons-Hawking term \cite{GH}  $S=S_{{\rm EH}}+S_{{\rm GH}}$  reads 
\eqn
S&=&-\frac{\sigma_{\perp}}{8\pi G}\int_{\M}\dd t\wedge\left\{-K_{\alpha}\wedge\dot\Sigma^\alpha+N\tilde W_{\alpha}\wedge\tilde e^{\alpha}+\sigma_\perp\hat Q\wedge K_{\beta}\wedge\tilde e^{\beta}\right.\nonumber\\
&&\phantom{-\frac{\sigma_{\perp}}{8\pi G}\int_{\M}\dd t\wedge\left\{\right.}\left.+\sigma_{\perp}q^{0\alpha}\tilde{\cal D}\Sigma_{\alpha}-N^{\alpha}\epsilon_{\alpha\beta\gamma}\tilde{\cal D}K^{\beta}\wedge\tilde e^{\gamma}\right\}\,,\label{lor.act}
\feqn
where we have introduced the oriented surface element $\Sigma^{\alpha}={}^{\tilde*}\tilde e^{\alpha}=\frac12\epsilon^{\alpha}{}_{\beta\gamma}\tilde e^{\beta}\wedge\tilde e^{\gamma}$, where ${}^{\tilde*}$ denotes the three-dimensional Hodge dual operator, the three dimensional Riemann 2-form $\rho_{\alpha}=\tilde\dd B_{\alpha}+\frac12\epsilon_{\alpha\beta\gamma}B^{\beta}\wedge B^{\gamma}$ and the three-dimensional exterior covariant derivative $\tilde{\cal D}^\alpha{}_\beta=\delta^\alpha{}_\beta\tilde\dd+\epsilon^\alpha{}_{\gamma\beta} B^\gamma\wedge$.

Using local Lorentz simmetry and diffeomorphism invariance we can gauge-fix $N=1$ and $N^\alpha=q^{0\alpha}=Q^\alpha=0$ \cite{MPT1}. Hence the equations describing any classical gravitational background in 4D are given by the zero torsion conditions 
\eq
K_\alpha\wedge\tilde e^\alpha=0\,,\qquad\tilde{\cal D}\tilde e^{\alpha}=0\,,\qquad\dot{\tilde e}^{\alpha}+K^{\alpha}=0\,,\label{lT.1}
\feq
and Einstein's equations
\eqn
\tilde W_{\alpha}\wedge\tilde e^{\alpha}=0\,,\qquad\epsilon_{\alpha\beta\gamma}\tilde{\cal D}K^\beta\wedge\tilde e^\gamma=0\,,\tilde W_{\alpha}+\epsilon_{\alpha\beta\gamma}\left(\dot K^\beta+\frac1{\ell^2}\tilde e^\beta\right)\wedge\tilde e^\gamma=0\,.\label{lE}
\feqn
An important role is played by the quantity 
\eqn
\label{tildeW}
\tilde W_{\alpha}=\rho_{\alpha}-\frac12\epsilon_{\alpha\beta\gamma}K^\beta\wedge K^\gamma+\frac1{\ell^{2}}\Sigma_{\alpha}\,,
\feqn
which is a component of the on-shell Weyl tensor $W^{ab}=R^{ab}+\sigma_\perp\ell^{-2}e^a\wedge e^b$, whose details can be read in Appendix \ref{app.Weyl}. Within our formalism and gauge-fixing the on-shell Weyl tensor reads
\eqn
\sigma_{\perp}W^{0\alpha}&=&\dd t\wedge\left(\dot K^\alpha+\frac1{\ell^2}\tilde e^\alpha\right)+\tilde{\cal D}K^\alpha\,,\label{l.Weyl.1}\\
W^\alpha&=&\frac{\sigma_\perp}2\epsilon^\alpha{}_{\beta\gamma}W^{\beta\gamma}=\dd t\wedge \dot B^\alpha+\tilde W^\alpha\,.\label{l.Weyl.2}
\feqn

We also review the Fefferman-Graham expansion for the various quantities in the 3+1-split formalism. The vielbein is expanded in powers of $e^{-t/\ell}$ as
\eq
\tilde e^\alpha=e^{t/\ell}E^\alpha(x)+e^{-t/\ell}\sum_{k=0} F^\alpha_{[k+2]}(x)e^{-kt/\ell}\,.\label{FG.vielbein}
\feq
where the finite term is missing due to the absence of a coupling to external sources. At the $t=+\infty$ boundary  $E^\alpha$ is a representative of a conformal class of vielbeins. Picking a particular defining function then we can refer to $E^\alpha$ as the {\it boundary vielbein}. Solving the equations \eqref{lT.1} we obtain
\eq
K^{\alpha}=-\frac1\ell e^{t/\ell}E^{\alpha}+\frac1\ell e^{-t/\ell}\sum_{k=0}(k+1)F^{\alpha}_{[k+2]}e^{-kt/\ell}\,.\label{FG.electric}
\feq
for the electric field and
\eq
\label{FG.magnetic}
B^{\alpha}=B_{[0]}+\sum_{k=2}B^{\alpha}_{[k]}e^{-kt/\ell}\,,
\feq
for the magnetic field. The first few coefficients of the latter are implicitly given by
\eqn
{\cal D}_{[0]}E^{\alpha}=0\,,\qquad{\cal D}_{[0]}F^{\alpha}_{[2]}+\epsilon^{\alpha}{}_{\beta\gamma}B^{\beta}_{[2]}\wedge E^{\gamma}=0\,,\qquad{\cal D}_{[0]}F^{\alpha}_{[3]}+\epsilon^{\alpha}{}_{\beta\gamma}B^{\beta}_{[3]}\wedge E^{\gamma}=0\,,\label{mag.FG}
\feqn
where ${\cal D}_{[0]}$ is the 3D covariant exterior derivative with respect to the boundary magnetic field, $B^{\alpha}_{[0]}$. Solving the first of \eqref{lT.1} and the Einstein's equations \eqref{lE}, the first few components of the expanded fields read:
\begin{description}
\item[$\{E^\alpha,B_{[0]}^\alpha\}${\bf.}] They represent geometric quantities  of the boundary being respectively the boundary vielbein and its torsionless spin connection.
\item[$\{F^\alpha_{[2]},B_{[2]}^\alpha\}${\bf.}] They represent respectively the boundary Schouten tensor
\begin{eqnarray*}
-\frac{2\sigma_{\perp}}{\ell^{2}}F^{\alpha}_{[2]}={}^{(3)}S^{\alpha}={\rm Ric}^{\alpha}-\frac{R}4E^{\alpha}\,.
\end{eqnarray*}
and the three-dimensional Hodge dual of the boundary Cotton-York tensor
\begin{eqnarray*}
B^\alpha_{[2]}=-\sigma_\perp{}^{\tilde*}{\cal D}_{[0]}F^\alpha_{[2]}=\frac{\ell^{2}}2{}^{\tilde*}C^{\alpha}\,.
\end{eqnarray*}
\item[$\{F^\alpha_{[3]}\}${\bf.}] This quantity is actually undetermined in the expansion. It  is symmetric $F^{\alpha}_{[3]}\wedge E_{\alpha}=0$, traceless $\epsilon_{\alpha\beta\gamma}F^{\alpha}_{[3]}\wedge E^{\beta}\wedge E^{\gamma}=0$ and obeys a conservation law $\epsilon_{\alpha\beta\gamma}{\cal D}_{[0]}F^{\beta}_{[3]}\wedge E^\gamma=0$. In the initial value formulation of gravity, this may be regarded as the {\it boundary initial velocity}, with $E^\alpha$ being the {\it boundary initial position} \cite{MPT1}. In a holographic setup (valid for $\sigma_\perp=1$), this determines the vacuum expectation value of the boundary energy momentum tensor.
\end{description}
Moreover, the components of the Weyl tensor read
\eqn
\dot K^{\alpha}+\frac1{\ell^{2}}\tilde e^{\alpha}&=&-\frac1{\ell^2}\sum_{k=0}^\infty[(k+2)^2-1]F^\alpha_{[k+3]}e^{-(k+2)t/\ell}\,,\label{FG.Weyl.expansion.1}\\
\tilde{\cal D}K^{\alpha}&=&-e^{-t/\ell}\frac2\ell\epsilon^{\alpha}{}_{\beta\gamma}B^{\beta}_{[2]}\wedge E^{\gamma}+{\cal O}\left(e^{-2t/\ell}\right)\,,\label{FG.Weyl.expansion.2}\\
\dot B^\alpha&=&-\frac1\ell\sum_{k=0}^\infty(k+2)B^{\alpha}_{[k+2]}e^{-(k+2)t/\ell}\,,\label{FG.Weyl.expansion.3}\\
\tilde W^{\alpha}&=&e^{-t/\ell}\frac3{\ell^{2}}\epsilon^{\alpha}{}_{\beta\gamma}F^{\beta}_{[3]}\wedge E^{\gamma}+{\cal O}\left(e^{-2t/\ell}\right)\,,\label{FG.Weyl.expansion.4}
\feqn
and a nice consequence of that is that the Weyl tensor as a whole vanishes at the $t\rightarrow+\infty $ boundary, $W^{ab}\Big|_{\partial\M}=0\,.$ We close this section giving the formula for computing the boundary stress tensor \cite{MPT1}, which is given by
\eq
\langle T_{ij}\rangle_{s}=E^\alpha{}_i\left({}^{\tilde*}\tau_\alpha\right)_j=\frac3{8\pi G\ell}F_{[3]\;ij}\,.\label{vev.stress.general}
\feq
where
\begin{eqnarray*}
\tau_\alpha\equiv\frac{\delta S_{{\rm ren.}}}{\delta E^\alpha}=\frac{3\sigma_\perp}{8\pi G\ell}\epsilon_{\alpha\beta\gamma}F^\beta_{[3]}\wedge E^\gamma=\frac{\sigma_\perp\ell}{8\pi G} \lim_{t\to+\infty}e^{t/\ell}\tilde W_{\alpha}\,,
\end{eqnarray*}
where $S_{{\rm ren.}}$ is the renormalized on-shell action.

\subsection{Self-duality with non-vanishing cosmological constant}
\label{sect.SD.lor}
In the following we introduce the notion of self-duality in gravity in the presence of a non-zero cosmological constant. Consider the Einstein-Hilbert action \eqref{EH.action} whose equations of motion read
\eq
\epsilon_{abcd}W^{ab}\wedge e^c=0\,,\qquad T^a=0\,.\label{self.eq.general}
\feq
If we take the exterior covariant derivative of the second of \eqref{self.eq.general} we end up with the Bianchi identity ${\cal D}T^{a}=R^a{}_b\wedge e^b=0$. This identity, which is actually an intergability condition for the equation $T^a=0$, can be modified to $W^{a}{}_{b}\wedge e^{b}=0$ by adding an identically vanishing term $\sigma_{\perp}\ell^{-2}e^{a}\wedge e_{b}\wedge e^{b}$. Therefore \eqref{self.eq.general} reads
\begin{eqnarray*}
{}^{\hat*}W^a{}_b\wedge e^b=0\,,\qquad T^a=0\quad\Rightarrow \quad W^a{}_b\wedge e^b=0\,,
\end{eqnarray*}
where ${}^{\hat*}$ is a \emph{tangent} Hodge dual operator, whose definition is given in \eqref{app.tang.hodge} in \ref{app.Weyl}. Therefore, if the torsion vanishes it is sufficient to set
\eq
W^{ab}=\pm{}^{\hat*}W^{ab}\,,\label{self.dual.general}
\feq
in order to have a solution to the equations of motion. This is the correct notion of self-duality in presence of a non-vanishing cosmological constant\footnote{Similar results were reported in \cite{Julia:2005ze}.}. In the case of Lorentzian signature \eqref{self.dual.general} has only the trivial solution
\begin{eqnarray*}
W^{ab}=0
\end{eqnarray*}
since ${}^{\hat*}{}^{\hat*}=-1$. Hence the only self-dual configurations are given by what we call {\it Weyl vacua}, a definition we are going to explain in the next section.

\section{4D Weyl Vacua {\it vs} 3D  gravitational Chern-Simons}
\label{sec.weyl.vacua}
Consider gravitational configurations  with vanishing torsion and Weyl tensor. These could be called {\it Weyl vacua} since they are related to flat connections of the conformal group. Indeed, as we discuss at the end of \ref{app.Weyl}, the  vielbein and the spin connection can be collected into a single $\mathfrak{g}$-valued connection
\eq
\label{Aconnection}
\A_{[4D]}=e^{a}P_{a}-\frac12\omega^{ab}J_{ab}\,,
\feq
with either $\mathfrak{g}={\rm so}(3,2)$ or $\mathfrak{g}={\rm so}(4,1)$ depending on whether $\sigma_{\perp}=1,-1$ respectively. $P_{a}$ are the generators of translations and $J_{ab}$ are the generators of Lorentz transformations obeying the following commutation rules
\eqn
\comm{J_{ab}}{J_{cd}}&=&\eta_{ad}J_{cb}-\eta_{cb}J_{ad}-\eta_{ac}J_{db}+\eta_{db}J_{ac}\,,\nonumber\\
\comm{J_{ab}}{P_{c}}&=&\eta_{ac}P_{b}-\eta_{bc}P_{a}\,,\\
\comm{P_{a}}{P_{b}}&=&-\Lambda J_{ab}\,.\nonumber
\feqn
The curvature of (\ref{Aconnection}) is given by
\begin{eqnarray*}
\F_{[4D]}=T^{a}P_{a}-\frac12W^{ab}J_{ab}\,,
\end{eqnarray*}
where $T^{a}$ is the torsion and $W^{ab}$ the on-shell Weyl tensor. Hence, torsionless and conformally flat metrics\footnote{Notice that actually, as explained in \ref{app.Weyl}, it is not possible to have torsionful conformally flat metrics by virtue of Bianchi identities.} are described by {\it flat} $\mathfrak{g}$-valued connections and thus can be considered as non-excited solutions. As we are going to show in the re\-main\-der of the section, these four-dimensional solutions are one-to-one related to classical configurations of a topological three-dimensional gauge theory: the gravitational Chern-Simons theory, or 3D conformal gravity \cite{Deser:1981wh,Horne:1988jf}. That three-dimensional Chern-Simons theory has an important role in the study of four-dimensional geometries with non-vanishing cosmological constant was also noticed in \cite{Cacciatori:2004rt,Cacciatori:2007vn} where it was shown that a subsector of BPS solutions to four-dimensional minimal gauged supergravity are described by a dimensionally reduced gravitational Chern-Simons theory. The hope is that the formulation we provided here could shed new light on finding geometric structures of gravitational configurations.

Considering the third equation of \eqref{lT.1} together with the vanishing of the first component of \eqref{l.Weyl.1} we get
\eq
\tilde e^\alpha=E^\alpha(x)e^{t/\ell}+F^\alpha_{[2]}(x)e^{-t/\ell}\,,\label{finite.FG.Weyl}
\feq
and hence the FG expansion of these solutions is finite \cite{SS}. According to the results reviewed in section \ref{resume}, at the  $t=+\infty$ boundary $E^{\alpha}$ is the boundary vielbein and $F^{\alpha}_{[2]}$ is  the boundary Schouten tensor. The vanishing of the $F^{\alpha}_{[3]}$ also implies that in a holographic setup (valid for $\sigma_\perp =1$) the vev of the boundary energy momentum tensor vanishes. Moreover, the vanishing of the first component of \eqref{l.Weyl.2} implies that the magnetic field is $t$-independent, $B^\alpha=B^\alpha_{[0]}(x)$. In particular, since the boundary Cotton tensor vanishes $B_{[2]}^\alpha=0$, the boudary metric must be conformally flat.

Gathering all together, the equations obeyed by the  fields $\{E^{\alpha}(x),\,F^{\alpha}_{[2]}(x),\,B^{\alpha}_{[0]}(x)\}$ are
\eqn
E_\alpha\wedge F_{[2]}^\alpha=0\,,\qquad{\cal D}_{[0]}E^{\alpha}={\cal D}_{[0]}F_{[2]}^{\alpha}=0\,,\nonumber\\
\rho_{[0]}^\alpha+\frac2{\ell^2}\epsilon_{\alpha\beta\gamma}F_{[2]}^\beta\wedge E^\gamma=0\,.\label{gathered}
\feqn
These  are {\it exactly} the equations of motion of a 3D Chern-Simons theory of the group $G={\rm SO}\left(3,2\right)$ for $\sigma_\perp=1$ or $G={\rm SO}\left(4,1\right)$ for $\sigma_\perp=-1$ whose solutions are given by conformally flat manifolds. To render more clear the relationship consider the Lie algebra $\mathfrak{g}$ of $G$ which is spanned by the conformal generators $\{\Pi_{\alpha},\,J_{\alpha},\,K_{\alpha},\,D\}$, satisfying the following commutation rules
\eqn
[\Pi_{\alpha},\Pi_{\beta}]&=&[K_{\alpha},K_{\beta}]=[J_{\alpha},D]=0\,,\nonumber\\
\left[D,\Pi_{\alpha}\right]&=&-\Pi_{\alpha}\,,\qquad[D,K_{\alpha}]=K_{\alpha}\,,\nonumber\\
\left[J_{\alpha},\Pi_{\beta}\right]&=&\sigma_{\perp}\epsilon_{\alpha\beta\gamma}\Pi^{\gamma}\,,\qquad[J_{\alpha},K_{\beta}]=\sigma_{\perp}\epsilon_{\alpha\beta\gamma}K^{\gamma}\,,\\
\left[\Pi_{\alpha},K_{\beta}\right]&=&-\epsilon_{\alpha\beta\gamma}J^{\gamma}+\eta_{\alpha\beta}D\,,\nonumber\\
\left[J_{\alpha},J_{\beta}\right]&=&\sigma_{\perp}\epsilon_{\alpha\beta\gamma}J^{\gamma}\,.\nonumber
\feqn
Decompose the $\mathfrak{g}$-valued connection $\A_{[3D]}$ in the following way
\begin{eqnarray*}
\A_{[3D]}=E^{\alpha}\Pi_{\alpha}+\sigma_{\perp}B_{[0]}^{\alpha}J_{\alpha}+\kappa F_{[2]}^{\alpha}K_{\alpha}+\phi D\,,
\end{eqnarray*}
where $\kappa$ is a real constant. The curvature of such a connection is given by
\eqn
\F_{[3D]}&=&\left({\cal D}_{[0]}E^{\alpha}-\phi\wedge E^{\alpha}\right)\Pi_{\alpha}+\sigma_{\perp}\left(\rho_{[0]}^\alpha-\sigma_{\perp}\kappa\epsilon^{\alpha}{}_{\beta\gamma}F_{[2]}^{\beta}\wedge E^{\gamma}\right)J_{\alpha}\nonumber\\
&&+\kappa\left({\cal D}_{[0]}F_{[2]}^{\alpha}+\phi\wedge F_{[2]}^{\alpha}\right)K_{\alpha}+\left(\tilde\dd\phi+\kappa E^{\alpha}\wedge F_{[2]\;\alpha}\right)D\,.\label{CS.curvature}
\feqn
Classical configurations of the Chern-Simons theory are trivial, since they are given by flat connections, say all the $\A_{[3D]}$ such that $\F_{[3D]}=0$. Picking $\kappa=-2\sigma_{\perp}/\ell^{2}$ and choosing the sector $\phi=0$ of the theory, we see that the system of equations we are left with is exactly the same as \eqref{gathered} which describes four-dimensional Weyl vacua. Moreover, since we have started with a gravitational theory in the bulk, $E^{\alpha}{}_{i}$ is an invertible matrix and it can be interpreted as the boundary vielbein. In this case, the field $\phi$ can always be gauge-fixed to zero \cite{Horne:1988jf}\footnote{An interesting generalization to this is studied in \cite{Cacciatori:2005wz} where the field $\phi$ is kept giving rise to Weyl structures in three dimensions.}. Therefore we conclude that the boundary data of all Weyl vacua of four dimensional gravity are described by the classical solutions of the three-dimensional gravitational Chern-Simons theory. In fact, we can go further and show that the 4D Weyl vacua and the classical configurations of the 3D gravitational Chern-Simons  are described by gauge-equivalent $\mathfrak{g}$-valued connections. The gauge transformation is given by the finite dilatation $g_{t}=e^{-tD/\ell}$ as 
\eqn
\A_{[4D]}&=&\dd tP_{0}+e^{t/\ell}E^\alpha\left(P_{\alpha}+\frac{\sigma_{\perp}}\ell J_{0\alpha}\right)+\sigma_{\perp}B_{[0]}^{\alpha}J_{\alpha}+e^{-t/\ell}F^\alpha_{[2]}\left(P_{\alpha}-\frac{\sigma_{\perp}}\ell J_{0\alpha}\right)\nonumber\\
&=&g_{t}\A_{[3D]}g_{t}^{-1}+g_{t}\dd g_{t}^{-1}
\feqn
where the generators are identified as
\begin{eqnarray*}
\Pi_{\alpha}=P_{\alpha}+\frac{\sigma_{\perp}}\ell J_{0\alpha}\,,\quad \kappa K_{\alpha}=P_{\alpha}-\frac{\sigma_{\perp}}\ell J_{0\alpha}\quad{\rm and}\quad D=\ell P_{0}\,.
\end{eqnarray*} 
Furthermore, we note that had we have chosen to consider the $t=-\infty$ boundary we would have  interpreted $F^{\alpha}{}_{[2]}$ as the boundary vielbein, and hence this, instead of $E^\alpha$, has to be invertible. This freedom in choosing the boundary vielbein has a precise interpretation from the 3D CS side: it correspond to exchanging the roles of translations $\Pi_{\alpha}$ and special conformal transformations $K_{\alpha}$. In CFT this is a highly non-trivial transformation, since usually the CFT spectrum is built using conformal modules whose base are quasi-primary operators annihilated by $K^\alpha$. However, we see that this non-trivial boundary transformation has a simple bulk image, namely the sign flip of the transverse coordinate $t$. 

Finally we can give an explicit representation  for Weyl vacua. For the sake of clarity we  choose $E^{\alpha}$ to be invertible. Then there exists a frame in which $E^{\alpha}=e^{\varphi(x)}\dd x^{\alpha}$ for an arbitrary scalar function $\varphi(x)$. As a consequence $B_{[0]}^{\alpha}=-\sigma_{\perp}\epsilon^{\alpha\beta}{}_{\gamma}\bar\partial_{\beta}\varphi\dd x^{\gamma}$, where $\bar\partial_{\alpha}=\partial/\partial x^{\alpha}$, and the Schouten tensor reads
\begin{eqnarray*}
\kappa F_{[2]}^{\alpha}={\rm Ric}^\alpha-\frac{R}4E^{\alpha}=e^{-\varphi}\left[T^\alpha{}_\beta-\bar\partial^\alpha\bar\partial_\beta\varphi\right]\dd x^\beta\,,
\end{eqnarray*}
where $T^\alpha{}_\beta=\bar\partial^{\alpha}\varphi\bar\partial_{\beta}\varphi-\frac12\nabla\varphi^{2}\delta^\alpha{}_\beta$ with $\nabla\varphi^{2}=\bar\partial\varphi\cdot\bar\partial\varphi$. As a result, the 4D vielbein becomes 
\begin{eqnarray*}
\tilde e^{\alpha}=e^{\varphi(x)-t/\ell}\dd x^{\alpha}-\sigma_{\perp}\frac{\ell^{2}}2e^{-\varphi(x)+t/\ell}\left[T^\alpha{}_\beta-\bar\partial^\alpha\bar\partial_\beta\varphi\right]\dd x^\beta\,.
\end{eqnarray*}

\subsection{Weyl structures in the boundary}
The 3+1-split formalism gives us very good control on the geometrical data, hence we want to study what changes in the discussion we did if we perform a different gauge-fixing on the geometric quantities. We will see that this will allow for more general geometric structures on the boundary. In section \ref{resume} we set $e^{0}{}_{i}=0$ from the beginning. Let us study in the following how the FG expansion and the boundary geometry are modified by keeping $e^0{}_{i}\neq0$. Setting  $e^{0}=N\dd t+\varphi$, where $\varphi=\varphi_{i}\dd x^{i}$ is an arbitrary 1-form, the action \eqref{lor.act} is modified by the following term
\begin{eqnarray*}
S_\phi=\frac{\sigma_{\perp}}{8\pi G}\int_{\M}\dd t\wedge\varphi\wedge\left[\left(\dot B_\alpha-\tilde{\cal D}Q_\alpha-\sigma_\perp\epsilon_{\alpha\beta\gamma}q^{0\beta}K^\gamma\right)\wedge\tilde e^\alpha+N^\alpha\left(\tilde W_\alpha+\frac2{\ell^2}\Sigma_\alpha\right)\right]\,.
\end{eqnarray*}
Leaving the detailed calculations in \ref{phi.eom}, the vanishing of the torsion constraints are modified as 
\eq
\sigma_\perp K_\alpha\wedge\tilde e^\alpha+\tilde\dd\varphi=0\,,\quad \tilde T^\alpha+\varphi\wedge K^\alpha=0\,,\quad \dot{\tilde e}^\alpha+K^\alpha=0\,,\label{lT.phi}
\feq
while Einstein's equations now read
\eqn
\tilde W_\alpha\wedge\tilde e^\alpha=0\,,\qquad\epsilon_{\alpha\beta\gamma}\left(\tilde{\cal D}K^\beta+\frac1{\ell^2}\varphi\wedge\tilde e^\beta\right)\wedge\tilde e^\gamma+\varphi\wedge\tilde W_\alpha=0\,,\nonumber\\
\tilde W_\alpha+\epsilon_{\alpha\beta\gamma}\left(\dot K^\beta+\frac1{\ell^2}\tilde e^\beta\right)\wedge\tilde e^\gamma-\varphi\wedge\dot B_\alpha=0\,.\qquad\qquad\label{lE.phi}
\feqn
Moreover, from the equations of motion, it follows that the field $\varphi$ does not depend on $t$, $\dot\varphi=0$. The on-shell Weyl tensor reads
\eqn
\sigma_\perp W^{0\alpha}&=&\dd t\wedge\left(\dot K^\alpha+\frac1{\ell^2}\tilde e^\alpha\right)+\tilde{\cal D}K^\alpha+\frac1{\ell^2}\varphi\wedge\tilde e^\alpha\,,\label{Weyl.tensor.phi1}\\
W^{\alpha}&=&\dd t\wedge \dot B^\alpha+\tilde W^\alpha\,.\label{Weyl.tensor.phi2}
\feqn
Hence the 4D metric we have at hand has the following form
\eq
\dd s^2=\sigma_\perp\left(\dd t+\varphi\right)^2+\dd s_{(3)}^2\,,\label{twistedspace}
\feq
where $\dd s_{(3)}^2=\tilde e^\alpha\tilde e_\alpha$ is the metric of the slice. The main difference from the previous situation is that the second of \eqref{lT.phi} introduces a new structure on the slice $\Sigma_t$: if we define a new connection $\hat\omega^{\alpha\beta}\equiv-\epsilon^{\alpha\beta\gamma}B_{\gamma}+\varphi K^{\alpha\beta}$ we still recover the vanishing of the three-dimensional torsion,
\begin{eqnarray*}
\hat T^\alpha\equiv\tilde\dd\tilde e^{\alpha}+\hat\omega^{\alpha}{}_{\beta}\wedge\tilde e^{\beta}=0\,,
\end{eqnarray*}
however the connection ceases to take values in the Lorentz algebra since $K^{\alpha\beta}$ has in general a non-vanishing symmetric part. This introduces {\it nonmetricity} in the boundary. Nonmetricity is a measure for the violation of local Lorentz invariance \cite{Hehl:1994ue}, which has become fashionable during the last years\footnote{We mention that, for example, nonmetricity (as well as torsion) have applications in the theory of defects in crystals, where they are interpreted as densities of point defects (line defects in case of torsion) \cite{Kroner:1990}.}. Therefore nonmetricity introduces a sort of pathological behaviour from the 3D point of view, but in 4D it is cured by a nontrivial immersion, described by \eqref{twistedspace}, of the boundary towards the inner of the bulk. We stress that $\varphi$ can always be set to zero by a gauge transformation, nevertheless its presence allows for having more general geometric structures on the boundary which however still generate, via standard evolution, non-pathological four-dimensional geometries.

Next we study Weyl vacua in this more general gauge fixing. From \eqref{lT.phi} and the vanishing of the Weyl tensor \eqref{Weyl.tensor.phi1}--\eqref{Weyl.tensor.phi2} it is clear that the vielbein $\tilde e^{\alpha}$ has the same finite expansion \eqref{finite.FG.Weyl} as before
\begin{eqnarray*}
\tilde e^{\alpha}=E^\alpha(x)e^{t/\ell}+F^\alpha_{[2]}(x)e^{-t/\ell}\,.
\end{eqnarray*}
The magnetic field is $t$-independent $B^{\alpha}=B^{\alpha}_{[0]}(x)$. The two fields $E^{\alpha}$ and $F^{\alpha}_{[2]}$ now obey to the following slightly modified equations
\begin{eqnarray*}
-\frac{2\sigma_{\perp}}\ell E_{\alpha}\wedge F^{\alpha}_{[2]}+\tilde\dd\varphi=0\,,&\qquad&{\cal D}_{[0]}E^{\alpha}-\frac{\varphi}\ell\wedge E^{\alpha}=0\,,\nonumber\\
\rho_{\alpha}+\frac2{\ell^{2}}\epsilon_{\alpha\beta\gamma}F_{[2]}^{\beta}\wedge E^{\gamma}=0\,,&\qquad&{\cal D}_{[0]}F_{[2]}^{\alpha}+\frac{\varphi}\ell\wedge F_{[2]}^{\alpha}=0\,.
\end{eqnarray*}
Notice now that by setting $\varphi/\ell=\phi$ these equations are exactly the conditions required for the vanishing of the connection $\F_{[3D]}$ \eqref{CS.curvature}. Hence Weyl vacua are still in one-to-one correspondence with classical configurations of the three-dimensional gravitational Chern-Simons theory. The new field $\varphi$ plays the role of the missing piece in the previous discussion: the component of the gauge field along dilatations $D$. Again we have that
\eqn
\A_{[4D]}=g_{t}\A_{[3D]}g_{t}^{-1}+g_{t}\dd g_{t}^{-1}\,,
\feqn
with the same transformation as before, but now $\left.\A_{[3D]}\right|_{D}=\varphi/\ell\neq 0$. 3D geometries with non-vanishing dilatational field were studied in detail in \cite{Cacciatori:2005wz} and they describe Weyl structures on conformally flat manifolds. Although they involve nonmetricity, they evolve into a non-pathological bulk geometry by the {\it twisted} evolution described in \eqref{twistedspace}.

\section{Euclidean signature and self-dual solutions}
Now we turn our attention to the notion of self-duality having in mind to study its consequences for holography. To do that we consider Euclidean signature and review below its salient differences with Lorentizan signature considered up to now. Using $\eta_{ab}={\rm diag}(+,+,+,+)$ we consider the usual 3+1-split for the vielbein, given by $e^0=N\dd t$ and $e^\alpha=N^\alpha\dd t+\tilde e^\alpha$, and accordingly for the spin connection $\omega^{0\alpha}=q^{0\alpha}\dd t+K^\alpha$, $\omega^{\alpha\beta}=-\epsilon^{\alpha\beta\gamma}\left(Q_\gamma\dd t+B_\gamma\right)$. We take the sum of the Einstein-Hilbert action plus the Gibbons-Hawking term that now reads
\eqn
S&=&-\frac1{8\pi G}\int_{\M}\dd t\wedge\left[-K_{\alpha}\wedge\dot\Sigma^\alpha-N\tilde W_{\alpha}\wedge\tilde e^{\alpha}-\hat Q\wedge K_{\beta}\wedge\tilde e^{\beta}\right.\nonumber\\
&&\phantom{-\frac1{8\pi G}\int_{\M}\dd t\wedge\left\{\right.}\left.+q^{0\alpha}\tilde{\cal D}\Sigma_{\alpha}-N^{\alpha}\epsilon_{\alpha\beta\gamma}\tilde{\cal D}K^{\beta}\wedge\tilde e^{\gamma}\right]\label{eucl.act}\,,
\feqn
where $\tilde W_{\alpha}=\rho_\alpha+\frac12\epsilon_{\alpha\beta\gamma}K^\beta\wedge K^\gamma-\Lambda\Sigma_{\alpha}$ is a component of the on-shell Weyl tensor. As before the only dynamical fields are the vielbein $\tilde e^\alpha$ and the electric field $K^\alpha$ and they are conjugate to each other. The other non-dynamical fields $\{N,N^\alpha,q^{0\alpha},Q^\alpha,B^\alpha\}$ can be considered as Lagrange multipliers. Exactly as in the Lorentzian case it is possible to fix some of them by virtue of the local symmetries of the theory, and hence we set $N=1$ and $N^{\alpha}=q^{0\alpha}=Q^{\alpha}=0$. Not all the symmetry has been used to provide such a gauge-fixing: in particular we are left with a residual $\widetilde{{\rm SO}(3)}$ group of transformations, which is the set of all three-dimensional rotations $\tilde g\in{\rm SO}(3)$ not depending on the $t$-coordinate. Within our gauge-fixing the whole Weyl tensor reads
\begin{eqnarray*}
W^{0\alpha}&=&\dd t\wedge\left(\dot K^\alpha+\Lambda\tilde e^\alpha\right)+\tilde{\cal D}K^\alpha\,,\\
W^\alpha&=&-\frac12\epsilon^\alpha{}_{\beta\gamma}W^{\beta\gamma}=\dd t\wedge \dot B^\alpha+\tilde W^\alpha\,.
\end{eqnarray*}
Varying the action we are left with
\eq
K_\alpha\wedge\tilde e^\alpha=0\,,\qquad\tilde{\cal D}\tilde e^{\alpha}=0\,,\qquad\dot{\tilde e}^{\alpha}+K^{\alpha}=0\,,\label{eT.1}
\feq
which state the vanishing of the torsion, plus Einstein's equations
\eqn
\tilde W_{\alpha}\wedge\tilde e^{\alpha}=0\,,\qquad\epsilon_{\alpha\beta\gamma}\tilde{\cal D}K^\beta\wedge\tilde e^\gamma=0\,,\qquad-\tilde W_{\alpha}+\epsilon_{\alpha\beta\gamma}\left(\dot K^\beta+\Lambda \tilde e^\beta\right)\wedge\tilde e^\gamma=0\,.\label{eE}
\feqn

\subsection{Simple self-dual solutions}
By design, gravity in the 3+1-split formalism resembles 4D Yang-Mills. Therefore, we expect that some of the salient properties of the latter will be shared by gravity as well.  For example, (see \ref{app.YM.inst} for a review), Euclidean YM configurations with equal electric and magnetic fields solve the equations of motion. We may ask then whether the corresponding condition provides gravitational solutions as well. Explicitly, if we set 
\begin{eqnarray*}
K^\alpha=\mp B^\alpha\,,
\end{eqnarray*}
we see that the first of \eqref{eE} necessarily implies that the cosmological constant must vanish $\Lambda=0$. The dynamical equations \eqref{eE} are automatically satisfied and hence we are only left with the vanishing of the torsion constraints which read 
\eq
K_\alpha\wedge\tilde e^\alpha=0\,,\quad\dot{\tilde e}^\alpha+K^\alpha=0\,,\quad
\tilde\dd\tilde e^\alpha\mp\epsilon^\alpha{}_{\beta\gamma}K^\beta\wedge\tilde e^\gamma=0\,.\label{eT.self.naive}
\feq
Plugging the second equation of \eqref{eT.self.naive} into the remaining two we end up with the following two equations for the vielbein
\eq
\dot{\tilde e}_\alpha\wedge\tilde e^\alpha=0\,,\qquad\dot\Sigma^\alpha=\mp\tilde\dd\tilde e^\alpha\,,\label{e-self}
\feq
which describe all self-dual solutions to Euclidean gravity without cosmological constant.

The system \eqref{e-self} can be simply solved providing known results. Consider for example the following ansatz for a cylindrically symmetric vielbein
\begin{eqnarray*}
\tilde e^i=r(t)\sigma^i\qquad {\rm for\;}i=1,2\,,\qquad \tilde e^3=r(t)G(t)\sigma^3\,,
\end{eqnarray*}
where $\sigma^\alpha$ are the ${\rm SU}(2)$-left-invariant 1-forms satisfying $\tilde\dd\sigma^\alpha=\epsilon^\alpha{}_{\beta\gamma}\sigma^\beta\wedge\sigma^\gamma$. The function $G(t)$ introduces an anisotropy in the spacetime metric. The electric field thus reads
\begin{eqnarray*}
K^i=-\dot r\sigma^i\,,\qquad K^3=-\dot{\left(rG\right)}\;\sigma^3
\end{eqnarray*}
and hence the ansatz satisfyies automatically the first condition of \eqref{e-self}. The latter therefore provides the constraint $\dot r=\mp G$ together with the following equation
\eq
r^2\dd(G^2)=2\left(1-G^2\right)\dd(r^2)\,.\label{EG-Han-eq}
\feq
One simple solution to this equation is given by $|G|=1$ (for definiteness take $G=1$). The solution becomes isotropic: the electric field is given by $K^\alpha=\pm\sigma^\alpha$ and since $r=\mp t$ up to a constant shift the metric can be written as
\begin{eqnarray*}
\dd s^2=\dd r^2+r^2\sigma^\alpha\sigma_\alpha\,,
\end{eqnarray*}
which describes $\mathbb{R}^4$ in spherical coordinates. If we consider instead $G$ to be a proper function of $t$, equation \eqref{EG-Han-eq} can be easily integrated to give
\begin{eqnarray*}
G=\left[1-\frac{a}{r^4}\right]^{1/2}\,,
\end{eqnarray*}
where $a$ is a real parameter. The electric field now reads $K^i=\pm G\sigma^i$ and $K^3=\pm(2-G^2)\sigma^3$. Trading the coordinate $t$ for $r$, the metric reads
\begin{eqnarray*}
\dd s^2=\frac{\dd r^2}{G^2}+r^2\left[\sigma^i\sigma_i+G^2(\sigma^3)^2\right]\,,
\end{eqnarray*}
which is the Eguchi-Hanson instanton \cite{Eguchi:1978gw}, the prototype of self-dual solutions to euclidean gravity without cosmological constant.

\subsection{Self-duality with non-vanishing cosmological constant}
The results of the previous subsection show that in the presence of a non-zero cosmological constant the notion of self-duality is no more connected to simply having equal electric and magnetic fields. The general idea was already given in section \ref{sect.SD.lor} for the Lorentizan case. The same holds here since the equations of motion descending from the Einstein-Hilbert action read
\begin{eqnarray*}
{}^{\hat*}W^a{}_b\wedge e^b=0\,,\qquad T^a=0\quad{\rm implies}\quad W^a{}_b\wedge e^b=0\,,
\end{eqnarray*}
and hence if the torsion vanishes it is sufficient to set
\eqn
W^{ab}=\pm{}^{\hat*}W^{ab}\label{eucl.SD}
\feqn
in order to have a solution. The difference here stands in the fact that ${}^{\hat*}{}^{\hat *}=+1$ due to the  Euclidean signature, and hence \eqref{eucl.SD} can have non-trivial solutions. In our formalism the self-duality condition
\begin{eqnarray*}
W^{0\alpha}=\pm\frac12\epsilon^\alpha{}_{\beta\gamma}W^{\beta\gamma}=\mp W^{\alpha}
\end{eqnarray*}
reads
\eqn
\left(K^\alpha\pm B^\alpha\right)\dot{}=-\Lambda\tilde e^\alpha\,,\label{Weyl-sf1}\\
\tilde\dd\left(K^\alpha\pm B^\alpha\right)\pm\frac12\epsilon^\alpha{}_{\beta\gamma}\left(K^\beta\pm B^\beta\right)\wedge\left(K^\gamma\pm B^\gamma\right)=\pm\Lambda\Sigma^\alpha\,.\label{Weyl-sf2}
\feqn

It is possible to re-write the system of equations describing self-dual solutions with non-vanishing cosmological constant in a nice way, making more evident the geometric structure of gravitational instantons. Remembering that the gauge-fixing we provided leaves a residual $\widetilde{{\rm SO}(3)}$ gauge freedom, we define a new ${\rm so}(3)$-valued connection
\begin{eqnarray*}
\pm\Omega^\alpha\equiv K^\alpha\pm B^\alpha\,,
\end{eqnarray*}
and hence eqns. \eqref{Weyl-sf1} and \eqref{Weyl-sf2} can be written in the following way
\eq
\dot\Omega^\alpha=\mp\Lambda\tilde e^\alpha\,,\qquad\rho^\alpha(\Omega)=\Lambda\Sigma^\alpha\,,\label{Weyl-sf2.geom}
\feq
where $\rho^\alpha(\Omega)\equiv\tilde\dd\Omega^\alpha+\frac12\epsilon^\alpha{}_{\beta\gamma}\Omega^\beta\wedge\Omega^\gamma$ is the curvature of $\Omega^\alpha$. Due to the vanishing of the torsion, $K^{\alpha}=-\dot{\tilde e}^{\alpha}$ and hence $B^{\alpha}=\Omega^{\alpha}\pm\dot{\tilde e}^{\alpha}$, hence the constraints we are left with are
\begin{eqnarray*}
\dot{\tilde e}_\alpha\wedge\tilde e^{\alpha}=0\,,\qquad\dot\Sigma^\alpha=\mp {\cal D}_\Omega\tilde e^\alpha\,.
\end{eqnarray*}
the latter being a ${\rm SO}(3)$-covariantization of equation \eqref{e-self}. Hence it is clear that turning on the cosmological constant means turning on a non trivial ${\rm so}(3)$-connection. To strengthen this statement set $\Lambda=0$, therefore \eqref{Weyl-sf2.geom} simplifies
\eqn
\dot\Omega^\alpha=0\,,\qquad\rho^\alpha(\Omega)=0\,,\label{Weyl-sf2.geom.nocosm}
\feqn
stating that the connection $\Omega^{\alpha}$ does not depend on $t$ and it is flat. Using the residual $\widetilde{{\rm SO}(3)}$ symmetry we can set it to zero $\Omega^{\alpha}=0$. 
As before, let us study some simple examples. Consider first an isotropic ansatz, given by $\tilde e^\alpha=r(t)\sigma^\alpha$. The electric and the magnetic fields are simply given by $
K^\alpha=-{\dot r}\sigma^\alpha$ and $B^\alpha=-\sigma^\alpha$ and hence the connection $\Omega^\alpha$ is isotropic since $\Omega^\alpha=-(1\pm \dot r)\sigma^\alpha$. Equations \eqref{Weyl-sf2.geom} simply give that $\ddot r=\Lambda r$ and $\dot r^{2}=1+\Lambda r^{2}$. Therefore trading the coordinate $t$ for $r$ the metrics can written in the following way
\begin{eqnarray*}
\dd s^2=\frac{\dd r^2}{1+\Lambda r^2}+r^2\sigma^\alpha\sigma_\alpha\,.
\end{eqnarray*}
which describes the metric of euclidean dS or AdS spacetimes, depending on the sign of the parameter $\Lambda$.

Consider now an Eguchi-Hanson-like ansatz, $\tilde e^i=r(t)f(t)\sigma^i$ and $\tilde e^3=r(t)g(t)\sigma^3$. As a result the electric and the magnetic fields read
\begin{eqnarray*}
K^i&=&-(rf)\dot{}\sigma^i\,,\qquad K^3=-(rg)\dot{}\sigma^3\,,\nonumber\\
B^i&=&-\frac{g}f\sigma^i\,,\qquad B^3=-\left(2-\frac{g^{2}}{f^{2}}\right)\sigma^3\,.
\end{eqnarray*}
Eqns. \eqref{Weyl-sf1}--\eqref{Weyl-sf2} with the plus sign are hence solved by picking
\begin{eqnarray*}
\dot r=f^{-2}=g^{-1}=1-\frac\Lambda2r^{2}\,,
\end{eqnarray*}
and the solution
\begin{eqnarray*}
\dd s^{2}=\frac{\dd r^{2}+r^{2}(\sigma^{3})^{2}}{(1-\Lambda r^{2}/2)^{2}}+\frac{r^{2}\sigma^{i}\sigma_{i}}{1-\Lambda r^{2}/2}
\end{eqnarray*}
is the well-know Fubini-Study metric on $P_{2}(\mathbb{C})$ \cite{Gibbons:1978zy}, which is an example of a self-dual configuration having  both a nut (at $r\sim0$) and a bolt (at $r\sim\infty$) removable singularities \cite{Gibbons:1979xm,Eguchi:1978gw}.

\section{Holography of self-dual configurations and the gravitational Chern-Simons}
To make connection with holography in the case of Euclidean signature we choose a negative cosmological constant, $\Lambda=1/\ell^{2}$. The vielbein, electric and magnetic fields are given by the same expressions as in the Lorentzian case, namely \eqref{FG.vielbein}, \eqref{FG.electric} and \eqref{FG.magnetic}. With  $E^{\alpha}$ being the boundary vielbein, the leading order component in the FG expansion of the magnetic field $B^{\alpha}_{[0]}$ is the boundary torsionless connection associated to it. The components \eqref{FG.Weyl.expansion.1}--\eqref{FG.Weyl.expansion.3} of the Weyl tensor are also the same here, however \eqref{FG.Weyl.expansion.4} is modified and now reads 
\eqn
\tilde W^{\alpha}&=&\rho^{\alpha}_{[0]}-\frac2{\ell^{2}}\epsilon^{\alpha}{}_{\beta\gamma}F^{\beta}_{[2]}\wedge E^{\gamma}-e^{-t/\ell}\frac3{\ell^{2}}\epsilon^{\alpha}{}_{\beta\gamma}F^{\beta}_{[3]}\wedge E^{\gamma}\nonumber\\
&&+e^{-2t/\ell}\left[{\cal D}_{[0]}B^\alpha_{[2]}-\frac4{\ell^2}\epsilon^\alpha{}_{\beta\gamma}F^\beta_{[4]}\wedge E^\gamma\right]+{\cal O}\left(e^{-3t/\ell}\right)\,.\label{FG.eucl.Weyl.4}
\feqn
The $F^{\alpha}_{[3]}$ and $F^{\alpha}_{[4]}$ components of the vielbein and the $B^{\alpha}_{[2]}$ and $B^{\alpha}_{[3]}$ components of the magnetic field are still traceless, symmetric matrices, and the three-dimensional component of the on-shell Weyl tensor reads
\begin{eqnarray*}
\tilde W^\alpha=-e^{-t/\ell}\frac3{\ell^{2}}\epsilon^{\alpha}{}_{\beta\gamma}F^{\beta}_{[3]}\wedge E^{\gamma}-e^{-2t/\ell}\frac8{\ell^2}\epsilon^\alpha{}_{\beta\gamma}F^\beta_{[4]}\wedge E^\gamma+{\cal O}\left(e^{-3t/\ell}\right)
\end{eqnarray*}
since $F^{\alpha}_{[2]}$ is still the proportional to the boundary Schouten tensor, $F^{\alpha}_{[2]}=-\frac{\ell^{2}}2{}^{(3)}S^{\alpha}$, and hence $B^{\alpha}_{[2]}$ is still proportional to the Hodge dual of the Cotton-York tensor,  $B^{\alpha}_{[2]}=-\frac{\ell^{2}}2{}^{\tilde*}C^{\alpha}$. Crucially, the divergence subtraction procedure, which is equivalent to setting up the initial value problem at the boundary \cite{MPT1}, can be copied from the Lorentzian case leading to the following result
\begin{eqnarray*}
\tau_{\alpha}=\frac{\left.\delta S_{{\rm ren.}}\right|_{\rm OS}}{\delta E^{\alpha}}=\frac3{8\pi G\ell}\epsilon_{\alpha\beta\gamma}F^{\beta}_{[3]}\wedge E^{\gamma}\,,
\end{eqnarray*}
and hence the vacuum expectation value of the energy momentum  tensor of the (Euclidean) boundary conformal field theory  is given by
\eq
\langle T_{ij}\rangle_{s}=E^{\alpha}{}_{i}\left({}^{\tilde*}\tau_{\alpha}\right)_{j}=-\frac3{8\pi G\ell}F_{[3]\;ij}\,.\label{vev.stress.eucl}
\feq

Now we are in a position to understand the consequence of self-duality in holography. By virtue of \eqref{FG.Weyl.expansion.1} and \eqref{FG.Weyl.expansion.3} we see that the self-duality conditions (\ref{Weyl-sf1}) and (\ref{Weyl-sf2}) provide algebraic relationships between the components of the FG expansion of the electric and magnetic fields. Explicitly we have
\eq
\label{SDcondition}
\frac1\ell\left[(k+2)^{2}-1\right]F^{\alpha}_{[k+3]}=\mp(k+2)B^{\alpha}_{[k+2]}\,.
\feq
For $k=0$ (\ref{SDcondition}) is an algebraic relationship between the {\it boundary initial velocity} $F^{\alpha}_{[3]}$ and the subleading component in the FG expansion of the magnetic field $B^{\alpha}{}_{[2]}$, which in turn is related to the Cotton-York tensor of the {\it boundary initial position}. Explicitly we have 
\begin{eqnarray*}
F^{\alpha}_{[3]}=\mp\frac23\ell B^{\alpha}_{[2]}=\pm\frac{\ell^{3}}3{}^{\tilde*}C^{\alpha}\,,
\end{eqnarray*}
and hence the vacuum expectation value of the boundary energy momentum tensor \eqref{vev.stress.eucl} is given by
\eq
\label{TCotton}
\langle T_{ij}\rangle=\mp\frac{\ell^{2}}{8\pi G}{}^{\tilde*}C_{ij}\,.
\feq
This result has been announced for the first time in \cite{dHPConf} and will be studied from a different perspective in the forthcoming work \cite{dHP2}.

Our results show that from the initial value formulation point of view, self-dual solutions to Euclidean gravity in four dimensions are associated to conformal invariant initial data. They are given by the evolution towards the bulk of boundary conformal classes, with initial velocity given by their  Cotton-York tensor, the only natural conformal tensor in three dimensions \cite{graham-2004}. Hence, self-dual metrics are uniquely determined by giving as initial data the conformal invariant pair $\{[E^{\alpha}],{}^{\tilde*}C^{\alpha}\}$, where $[E^{\alpha}]$ denotes the conformal class having $E^{\alpha}$ as representative, in agreement with the results contained in \cite{fefferman-2007}.

From the holographic point of view, we have reached a remarkable conclusion: the {\it exact} boundary generating functional of self-dual configurations is the three-dimensional gravitational Chern-Simons theory, for the functional derivative of the latter with respect to the boundary vielbein (or equivalently the boundary metric) gives the Cotton tensor.\footnote{The easiest way to see this is e.g. \cite{Hehl} to write down the three-dimensional CS action imposing the vanishing of the torsion by a Lagrange multiplier, the latter then turns out to be proportional to the Schouten tensor. The variation with respect to the vielbein gives the covariant derivative of the Largange multiplier and hence the Cotton tensor.}  Succinctly, the CFT dual to self-dual four-dimensional metrics is holographically described by the three-dimensional gravitational Chern-Simons theory. 

However, our result (\ref{TCotton}) accepts a different holographic interpretation. We can interpret the boundary veilbein as a {\it classical graviton} and then (\ref{TCotton}) tells us that the three-dimensional gravitational Chern-Simons is (minus) the leading term in the {\it  effective action}  for the boundary theory describing self-dual configurations. To sketch how this may come about, consider the generating functional of the 3D (Euclidean) gravitational CS theory 
\begin{eqnarray*}
e^{W[T]}=\int[{\cal D} h]e^{-S_{{\rm CS}}[h]+\int h_{ij}T^{ij}}\,,
\end{eqnarray*}
where $h_{ij}=E^{\alpha}{}_{i}E_{\alpha\,j}$ and $T_{ij}$ is an external matter source. If we called $h^{{\rm cl.}}=\langle h\rangle_{T}$, the effective action, which is the Legendre transformation of the generating functional
\begin{eqnarray*}
\Gamma[h^{{\rm cl.}}]=W[T]-\int h^{\rm cl.}_{ij}T^{ij}\,,
\end{eqnarray*}
would read
\eqn
\Gamma[h^{\rm cl.}]=-S_{{\rm CS}}[h^{\rm cl.}]-\frac12\log{\rm det}\left[\frac{\delta^{2} S_{{\rm CS}}}{\delta h\delta h}\right]\Big|_{h^{{\rm cl.}}}+\cdots\,.
\feqn
The possibility of such a modified holographic dictionary in the context of AdS$_4$/CFT$_3$ has appeared before for the case of scalar \cite{dHP1,dHPP}, vectors \cite{dHG} and recently gravity \cite{dHPConf,Marolf}. 

\section{Conclusions}
In this and the companion work \cite{MPT1} we  have presented a detailed analysis of gravity in the presence of a cosmological constant in the 3+1-split formalism having in mind applications to AdS$_4$/CFT$_3$. Here we focused on the issue of self-duality. In Lorentzian signature, the only self-dual configurations are the Weyl vacua and we have shown how they are holographically described by classical solutions of the three-dimensional gravitational Chern-Simons theory. In the case of Euclidean signature, the non-trivial bulk self-dual configurations are also related to the same three-dimensional Chern-Simons theory: the latter theory can either be viewed as the exact generating functional of the boundary CFT or the leading term in its effective action. These observations appear to be closely related to past work on quantum gravity\footnote{We thank Lee Smolin for bringing these works to our attention.}, in particular on its holographic formulation \cite{Smolin1,Smolin2,Smolin3}.

Our results emphasise once more the distinctive properties of AdS$_4$/CFT$_3$ correspondence. Namely, it appears to allow for an  {\it exact} holographic description of three-dimensional CFTs based on Chern-Simons models. In this sense, the recent interest in such 3D theories \cite{BL1,BL2,BL3,Gustavsson,ABJM} offers an exciting prospect. 

\section*{Acknowledgements}
We would like to thank D.~Klemm, G.~Kofinas, D.~Minic and R.~Olea for helpful discussions. A.~C.~P. would like to thank S.~de~Haro and R.~G.~Leigh for long term enlightening discussions and collaboration on similar ideas, and the latter for a critical reading of the manuscript. D.~M. and A.~C.~P. were partially supported by the INTERREG IIIA Program K2301.007. A.~C.~P. was partially supported by the European RTN Program MRTN-CT-2004-512194. G.~T. was partially supported by the Italian MIUR-PRIN contract.

\appendix

\section{Weyl's Conformal Tensor}
\label{app.Weyl}
Consider a four-dimensional manifold $\M$ endowed with  a metric structure described by a vielbein $e^a$ and a torsionless ${\rm so}(3,1)$-valued connection $\omega^{ab}$. The Riemann tensor, given explicitly as the curvature of the Lorentz connection, $R^{ab}=\dd\omega^{ab}+\omega^a{}_c\wedge\omega^{cb}$, can be decomposed into the following parts which are irreducible representations of the full Lorentz group
\eqn
R^{ab}=C^{ab}+E^{ab}+G^{ab}\,,\label{Weyl.dec.def}
\feqn
where
\eqn
E^{ab}=e^{[a}\wedge F^{b]}\,,\qquad G^{ab}=\frac{R}{12}e^a\wedge e^b\label{Weyl.dec.def2}
\feqn
being $F^a={\rm Ric}^a-\frac{R}4e^a$ the traceless part of the Ricci tensor $R^{a}{}_b$, ${\rm Ric}^a=R^a{}_be^b$ the Ricci 1-form and $R=e
_a\rfloor{\rm Ric}^a=R^a{}_a$ the scalar curvature. This decomposition defines the \emph{Weyl conformal tensor} $C^{ab}$: it is called ``conformal'' since its components do not change under conformal transformations. It is possible to define the Weyl tensor in any dimensions $D$ (actually for $D>3$) in the following way
\begin{eqnarray*}
C^{ab}\equiv R^{ab}-e^{a}\wedge S^{b}+e^b\wedge S^a\,,
\end{eqnarray*}
where
\begin{eqnarray*}
S^a\equiv\frac1{D-2}\left[{\rm Ric}^a-\frac{R}{2(D-1)}e^a\right]
\end{eqnarray*}
is the Schouten tensor. Besides the standard symmetries enjoyed by the Riemann tensor, the Weyl tensor has the additional feature to be completely traceless, $e_a\rfloor C^{ab}=0$, and hence in four dimensions it has ten independent components. In three dimensions it turns out that the Weyl tensor vanishes identically and thus the Riemann tensor is given entirely in terms of the Schouten tensor, 
\begin{eqnarray*}
{}^{(3)}R^{ab}=e^a\wedge {}^{(3)}S^b-e^b\wedge {}^{(3)}S^a\,,
\end{eqnarray*}
with
\begin{eqnarray*}
{}^{(3)}S^a={\rm Ric}^a-\frac{R}4e^a\,.
\end{eqnarray*}
Let us go back to the original definition (\ref{Weyl.dec.def}), given in the case of a four dimensional manifold. When we consider ${\rm so}(3,1)$-valued 2-forms $\Lambda^{ab}=\frac12\Lambda^{ab}{}_{cd}e^c\wedge e^d$, such as any term in (\ref{Weyl.dec.def}), we can deal with two different notions of Hodge duality: one concerning the flat, tangent indices
\eqn
{}^{\hat*}\Lambda^{ab}=\frac12\epsilon^{ab}{}_{a'b'}\Lambda^{a'b'}=\frac14\epsilon^{ab}{}_{a'b'}\Lambda^{a'b'}{}_{cd}e^c\wedge e^d\,,\label{app.tang.hodge}
\feqn
and one concerning curved, spacetime indices
\begin{eqnarray*}
{}^*\Lambda^{ab}=\frac12\Lambda^{ab}{}_{c'd'}{}^*\left(e^{c'}\wedge e^{d'}\right)=\frac14\Lambda^{ab}{}_{c'd'}\epsilon^{c'd'}{}_{cd}e^c\wedge e^d\,.
\end{eqnarray*}
The two notions, in general, have nothing to do with each other. But, from the definitions we gave in (\ref{Weyl.dec.def2}), it turns out that
\begin{eqnarray*}
{}^{\hat*}C^{ab}={}^*C^{ab}\,,\qquad {}^{\hat*}E^{ab}=-{}^*E^{ab}\,,\qquad{}^{\hat*}G^{ab}={}^*G^{ab}\,.
\end{eqnarray*}
If Einstein's equations hold, in absence of any source term, ${\rm Ric}^{a}=(R/2+3\Lambda)e^a$, the $E^{ab}$ component of the Weyl tensor vanishes and hence the on-shell Riemann tensor reads $R^{ab}=C^{ab}-\Lambda e^a\wedge e^b$, having the property ${}^{\hat*}R^{ab}={}^*R^{ab}$. So that the tensor $W^{ab}=R^{ab}+\Lambda e^a\wedge e^b$ we used throughout the paper can be reasonably called the \emph{on-shell Weyl tensor}.

This tensor has another interesting geometric interpretation. The fundamental fields in gravity, say the vielbein and the spin connection, can be assembled into a single Lie algebra-valued connection. For the case of four-dimensional gravity with a non vanishing cosmological constant (the case with vanishing cosmological constant can then be recovered by an Inonu-Wigner contraction) we consider the Lie group $G={\rm SO}\left(3,2\right)$ or $G={\rm SO}\left(4,1\right)$, depending on wheter $\sigma_\perp=\pm 1$ respectively, whose algebra $\mathfrak{g}$ is generated by the standard four-di\-men\-sio\-nal Poincar\'e generators, $P_a$ and $J_{ab}$ with $a,b=0,1,2,3$, with the introduction of a non-commutativity between translations
\begin{eqnarray*}
[P_a,P_b]=-\Lambda J_{ab}\,.
\end{eqnarray*}
Picking a $\mathfrak{g}$-valued connection $\A$, it is natural to interpret its components along generators as $\A=e^aP_a-\frac12\omega^{ab}J_{ab}$, where $e^a$ is the vielbein and $\omega^{ab}$ the spin connection. Its curvature $\F=\dd\A+\A\wedge\A$ can thus be written as $\F=T^aP_a-\frac12W^{ab}J_{ab}$, where $T^a$ is the standard definition for the torsion and $W^{ab}=R^{ab}+\Lambda e^a\wedge e^b$ precisely. So that $W^{ab}$ has a geometric interpretation: it is the component of the curvature of a $\mathfrak{g}$-valued connection along Lorentz transformations.

Within this last context one should pay special attention to the Bianchi identities, since the $G$-covariant exterior derivative is different from the simple Lorentz-covariant one due to the presence of the translations. In particular the Bianchi identity reads $\nabla\F=0$, where $\nabla\F=\dd\F+\A\wedge\F-\F\wedge\A$ with, whose components read
\eqn
\nabla\F\Bigl|_{P}&=&{\cal D}T^{a}-W^{a}{}_{b}\wedge e^{b}=0\,,\nonumber\\
\nabla\F\Bigl|_{J}&=&{\cal D}W^{ab}+\Lambda e^{a}\wedge T^{b}-\Lambda e^{b}\wedge T^{a}=0\,,\label{Bianchi.W}
\feqn
where ${\cal D}$ is the Lorentz-covariant part of the full $\nabla$. An interesting fact is that it is not possible to have a configuration with vanishing $W^{ab}$ and non-vanishing torsion $T^{a}$, being the condition $W^{ab}=0$ even more restrictive than $R^{ab}=0$.

\section{Details of computations for non-vanishing $e^{0}{}_{i}$}
\label{phi.eom}
Let us give the details for computing the equations of motion when the field $e^{0}{}_{i}=\varphi$ ceases to vanish, and hence to compute the modifications to variations of the action \eqref{lor.act} when the following term is included
\begin{eqnarray*}
S_\phi=\frac{\sigma_{\perp}}{8\pi G}\int_{\M}\dd t\wedge\varphi\wedge\left[\left(\dot B_\alpha-\tilde{\cal D}Q_\alpha-\sigma_\perp\epsilon_{\alpha\beta\gamma}q^{0\beta}K^\gamma\right)\wedge\tilde e^\alpha+N^\alpha\left(\tilde W_\alpha+\frac2{\ell^2}\Sigma_\alpha\right)\right]\,.
\end{eqnarray*}
Let us call $\hat S=S+S_\phi$. The fields $N,\,N^{\alpha},\,q^{0\alpha},\,Q^{\alpha}$ still play the role of Lagrange multipliers and hence the constraints they provide are modified to, respectively
\eqn
\tilde W_\alpha\wedge\tilde e^{\alpha}=0\,,\label{lC.1.phi}\\
\epsilon_{\alpha\beta\gamma}\left(\tilde{\cal D}K^\beta+\frac1{\ell^2}\varphi\wedge\tilde e^\beta\right)\wedge\tilde e^\gamma+\varphi\wedge\tilde W_\alpha=0\,,\label{lC.2.phi}\\
\epsilon_{\alpha\beta\gamma}\left(\tilde T^{\beta}+\varphi\wedge K^\beta\right)\wedge\tilde e^{\gamma}=0\,,\label{lC.3.phi}\\
\tilde e_\alpha\wedge \left(\sigma_\perp K_\beta\wedge\tilde e^\beta+\tilde\dd\varphi\right)-\varphi\wedge\tilde T_\alpha=0\,,\label{lC.4.phi}
\feqn
where $\tilde T^{\alpha}=\tilde{\cal D}\tilde e^{\alpha}$. As in the standard case with vanishing $\varphi$ it is still possible to gauge-fix $N=1$ and $N^{\alpha}=q^{0\alpha}=Q^{\alpha}=0$. Hence the variations with respect to the magnetic field and the new field $\varphi$ read respectively
\eq
\tilde T_{\alpha}-\left(\varphi\wedge\tilde e_\alpha\right)\dot{}=0\,,\qquad \dot B_\alpha\wedge\tilde e^\alpha=0\,.\label{lC.5.phi}
\feq
Furthermore, the variations with respect to the dynamical fields, the electric field and the vielbein, give respectively
\eqn
\dot{\tilde e}^\alpha+K^\alpha=0\,,\label{lE.1.phi}\\
\tilde W_\alpha+\frac2{\ell^2}\Sigma_\alpha+\epsilon_{\alpha\beta\gamma}\tilde e^\beta\wedge\dot K^\gamma-\varphi\wedge\dot B_\alpha=0\,.\label{lE.2.phi}
\feqn
Taking the exterior multiplication of the first of \eqref{lC.5.phi} with $\epsilon_{\alpha\beta\gamma}\tilde e^{\gamma}$, by virtue of eqn. \eqref{lC.3.phi}, we obtain $\dot\varphi=0$. Hence the field $\varphi$ is a pure boundary object and it does not evolve into the bulk. A striking consequence of the presence of $\varphi$ is that the three-dimensional torsion $\tilde T^{\alpha}$ does not vanish in general, since from the first of \eqref{lC.5.phi}
\eq
\tilde T^\alpha=-\varphi\wedge K^\alpha\,,\label{lC.6.phi}
\feq
and hence eqn. \eqref{lC.3.phi} becomes redundant since eqn. \eqref{lC.6.phi} provides a much stronger condition. Finally eqn. \eqref{lC.4.phi} reduces to
\begin{eqnarray*}
\sigma_\perp K_\alpha\wedge\tilde e^\alpha+\tilde\dd\varphi=0\,,
\end{eqnarray*}
and hence the derivatives of $\varphi$ measure the failure of $K_\alpha$ to be symmetric in this new setup. Therefore the equations we are left with are given exactly by \eqref{lT.phi} and \eqref{lE.phi}.

\section{Euclidean Yang-Mills theory and self-dual solutions}
\label{app.YM.inst}
We want to develop the first order formalism for a generic YM theory for some Lie group $G$. Call $\A=\varphi\dd t+\tilde\A$ the $\mathfrak{g}$-valued connection and $\F=\dd t\wedge E+\tilde\F\;$ ($E$ is a 1-form such that $E(\partial_t)=0$ and $\tilde\F$ is a 2-form such that $\tilde\F(\partial_t,\bullet)=0$) a $\mathfrak{g}$-valued 2-form which, on-shell, shall give the curvature of the potential $\A$, say $\F=\dd\A+\A\wedge\A$. Pick a manifold $\M$, endowed with a metric structure $g$ providing the standard Hodge dual operator ${}^*$. Therefore we have for the field $\F$
\begin{eqnarray*}
{}^*\F=\dd t\wedge B+\widetilde{{}^*\F}\,,
\end{eqnarray*}
where
\begin{eqnarray*}
B_i=\sqrt{g}\epsilon_{ijk}\left(g^{jt}E^k+\frac12\tilde\F^{jk}\right)\,,\qquad\widetilde{{}^*\F}_{ij}=\sqrt{g}\epsilon_{ijk}\left(g^{tt}E^k-g^{tk}E^t+g^{tl}g^{km}\tilde\F_{lm}\right)\,,
\end{eqnarray*}
where $\epsilon_{ijk}=\epsilon_{tijk}$ are the three-dimensional Levi-Civita symbols. It is always possible to choose well-adapted coordinates in order to set $g_{tt}=1$ and $g_{ti}=0$. In this way the metric on $\M$ can be written as
\begin{eqnarray*}
\dd s^2=\dd t^2+h_{ij}(t,\vec x)\dd x^i\dd x^j\,,
\end{eqnarray*}
and hence the dual of $\F$ simplifies to
\begin{eqnarray*}
B={}^{\tilde*}\tilde\F\,,\qquad\widetilde{{}^*\F}={}^{\tilde*}E\,.
\end{eqnarray*}
Picking an Ad-invariant, symmetric, non-degenerate bilinear form $\langle\bullet,\bullet\rangle$ on the algebra, the action shall read
\begin{eqnarray*}
S&=&\int_{\M}-\frac12\langle\F\wedge{}^{*}\F\rangle+\langle\left(\dd\A+\A\wedge\A\right)\wedge{}^{*}\F\rangle\,,\nonumber\\
\phantom{S}&=&\int_{\M}\dd t\wedge\left[\langle\dot{\tilde \A}\wedge\widetilde{{}^*\F}\rangle-\frac12\left(\langle E\wedge\widetilde{{}^*\F}\rangle+\langle\tilde\F\wedge B\rangle\right)\right.\nonumber\\
&&\phantom{\int_{\M}\dd t\wedge}\left.\phantom{\frac12}+\langle\left(\tilde\dd\tilde\A+\tilde\A\wedge\tilde\A\right)\wedge B\rangle+\langle\varphi,\tilde\nabla\widetilde{{}^*\F}\rangle\right]-\int_{\M}\dd t\wedge\tilde\dd\langle\varphi,\widetilde{{}^*\F}\rangle\,,
\end{eqnarray*}
where the last term is actually a boundary term. Equivalently, if we performed the transformation to bring the metric in the preferred form, the action would read
\eqn
S&=&\int_{\M}\dd t\wedge\left[\langle\dot{\tilde \A}\wedge{}^{\tilde *}E\rangle-\frac12\left(\langle E\wedge{}^{\tilde *}E\rangle+\langle{}^{\tilde *}B\wedge B\rangle\right)\right.\nonumber\\
&&\phantom{\int_{\M}\dd t\wedge}\left.\phantom{\frac12}+\langle\left(\tilde\dd\tilde\A+\tilde\A\wedge\tilde\A\right)\wedge B\rangle+\langle\varphi,\tilde\nabla{}^{\tilde *}E\rangle\right]+\int_{\partial\M}\dd t\wedge\langle\varphi,{}^{\tilde*}E\rangle\,.
\feqn
It is easy, at this point, interpreting the fields. $\varphi$ plays the role of a Lagrange multiplier for the constraint $\tilde\nabla{}^{\tilde *}E=0$, the Gauss law, which is obtained by varying the action with respect to  $\varphi$ itself. The dynamical fields, conjugate to each other, are given by the potential $\tilde\A$ and the electric field $E$, while the magnetic field is some external field. The Lagrange multiplier can be fixed to zero by a gauge transformation, say a certain $g\in G$ such that $\varphi=g^{-1}\dot g$. Hence we are left with a residual gauge symmetry given by group elements $\tilde g\in G$ such that $\dot{\tilde g}=0$. Therefore, within this gauge fixing, the equations of motion read
\begin{eqnarray*}
\frac{\delta S}{\delta {}^{\tilde *}E}&=&\dot{\tilde{\A}}-E=0\,,\\
\frac{\delta S}{\delta\tilde\A}&=&-\left({}^{\tilde *}E\right)\dot{}+\tilde\nabla B=0\,,\\
\frac{\delta S}{\delta B}&=&\tilde\dd\tilde\A+\tilde\A\wedge\tilde\A-{}^{\tilde *}B=0\,,
\end{eqnarray*}
plus the Gauss law
\begin{eqnarray*}
\tilde\nabla{}^{\tilde *}E=0\,.
\end{eqnarray*}

\subsection{Self-dual solutions}
We want look for solutions enjoying $B=\pm E$. The equations thus read
\begin{eqnarray*}
E=\dot{\tilde{\A}}\,,\qquad{}^{\tilde *}E=\pm\left(\tilde\dd\tilde\A+\tilde\A\wedge\tilde\A\right)\,,\qquad{}^{\tilde *}\dot E=\pm\tilde\nabla E\,,\qquad\tilde\nabla{}^{\tilde *}E=0\,.
\end{eqnarray*}
Actually the Gauss law is equivalent to the Bianchi identity and thus it is automatically satisfied. The dynamical equation is moreover automatically satisfied if the following holds
\begin{eqnarray*}
\pm({}^{\tilde *}\tilde\A)\dot{}=\tilde\dd\tilde\A+\tilde\A\wedge\tilde\A\,,
\end{eqnarray*}
which is a first order condition on the connection. Notice that the equations are invariant under the residual gauge freedom. Consider the simple case where $G={\rm U}(1)^{n}$ (or its non-compact version): the connection is thus described by $n$ commuting 1-forms $\A^{I}$ and hence the equation for self-dual solutions reads
\begin{eqnarray*}
\pm({}^{\tilde *}\tilde\A^{I})\dot{}=\pm\tilde\dd\A^{I}\,.
\end{eqnarray*}
Consider now a Lie group $G$ whose algebra $\mathfrak{g}$ is given as the sum $\mathfrak{g}=\mathfrak{h}+\mathfrak{k}$ where $\mathfrak{h}$ is a subalgebra and $\mathfrak{k}$ is weakly reducible $\comm{\mathfrak{h}}{\mathfrak{k}}\subseteq\mathfrak{k}$ and symmetric $\comm{\mathfrak{k}}{\mathfrak{k}}\subseteq\mathfrak{h}$. Hence the generators of $\mathfrak{g}$ can be split into $\{T_{I}\}=\{T_{H},T_{K}\}$, where $T_{H}$ and $T_{K}$ span respectively $\mathfrak{h}$ and $\mathfrak{k}$, and the commutation rules can be schematically written as
\begin{eqnarray*}
\left[T_{H},T_{H'}\right]&=&f_{HH'}{}^{H''}T_{H''}\,,\nonumber\\
\left[T_{H},T_{K}\right]&=&f_{HK}{}^{K'}T_{K'}\,,\\
\left[T_{K},T_{K'}\right]&=&f_{KK'}{}^{H}T_{H}\,.\nonumber
\end{eqnarray*}
Hence the equation reads
\begin{eqnarray*}
\pm({}^{\tilde *}\tilde\A^{H})\dot{}&=&\pm\left(\tilde\F^{H}+\frac12f_{K'K''}{}^{H}\tilde\A^{K'}\wedge\tilde\A^{K''}\right)\,,\\
\pm({}^{\tilde *}\tilde\A^{K})\dot{}&=&\pm\tilde\nabla^{(H)}\tilde\A^{K}\,,
\end{eqnarray*}
where $\tilde\F^{H}=\tilde\dd\tilde\A^{H}+\frac12f_{H'H''}{}^{H}\tilde\A^{H'}\wedge\tilde\A^{H''}$ is the curvature of the connection $\A^H$, related to sublgebra $\mathfrak h$, and $\tilde\nabla^{(H)}\tilde\A^{K}=\tilde\dd\tilde\A^{K}+f_{HK'}{}^{K}\tilde\A^{H}\wedge\tilde\A^{K'}$ where $\tilde\nabla^{(H)}$ is the $H$-covariant exterior derivative, being $H$ the subgroup associated to the subalgebra $\mathfrak{h}$. For instance consider the groups ${\rm SO}(4)$ or ${\rm SO}(3,1)$ whose algebra is given by
\begin{eqnarray*}
\left[P_\alpha,P_\beta\right]&=&-\lambda\epsilon_{\alpha\beta}{}^\gamma J_\gamma\,,\nonumber\\
\left[J_\alpha,P_\beta\right]&=&\epsilon_{\alpha\beta}{}^\gamma P_\gamma\,,\\
\left[J_\alpha,J_\beta\right]&=&\epsilon_{\alpha\beta}{}^\gamma J_\gamma\,,\nonumber
\end{eqnarray*}
where $\alpha,\beta=0,1,2$, $\eta={\rm diag}(+,+,+)$ and $\lambda$ is a parameter whose sign is determined by the choice of the signature of the gauge group. If $\tilde\A=\tilde e^\alpha P_\alpha+\Omega^\alpha J_\alpha$ then the equations for instantons are given by
\begin{eqnarray*}
\left({}^{\tilde*}\Omega^\alpha\right)\dot{}&=&\pm\left(\tilde\dd\Omega^\alpha+\frac12\epsilon^\alpha{}_{\beta\gamma}\Omega^\beta\wedge\Omega^\gamma-\frac\lambda2\epsilon^\alpha{}_{\beta\gamma}\tilde e^\beta\wedge\tilde e^\gamma\right)\,,\\
\left({}^{\tilde*}\tilde e^\alpha\right)\dot{}&=&\pm\left(\tilde\dd\tilde e^\alpha+\epsilon^\alpha{}_{\beta\gamma}\Omega^\beta\wedge\tilde e^\gamma\right)\,.
\end{eqnarray*}


\end{document}